\documentclass[twocolumn,preprintnumbers,superscriptaddress]{revtex4-2}
\usepackage{hyperref}
\usepackage[utf8]{inputenc}
\usepackage[skip=10pt]{caption} 

\usepackage[table,xcdraw]{xcolor} % for table colors

\usepackage{enumitem}
\usepackage[nopatch]{microtype}
\usepackage{balance}
\usepackage{makecell}
\usepackage{mhchem}

\usepackage{soul}
\usepackage{url}
\usepackage{mathtools}
\usepackage{braket}
\usepackage{amsmath}
\usepackage{dsfont}
\usepackage[english]{babel}
\usepackage{graphicx} %package to manage images
\usepackage{tikz}
\usetikzlibrary{quantikz}

\usepackage{hyperref}
\usepackage{natbib} % this is probably into you nips_2018 format file
\usepackage{physics}

\usepackage{listings}
\usepackage{xcolor}
\usepackage{amssymb}

\usepackage{babel}

\newcommand{\LDIG}{
Lighthouse Disruptive Innovation Group, LLC
1 Broadway, 14th floor, Cambridge,
Middlesex County, Massachusetts 02142 (USA)
}

\newcommand{\LDIGEU}{
Lighthouse Disruptive Innovation Group Europe, SL.
Barcelona - Spain
}

\newcommand{\MIT}{
MIT Media Lab - City Science Group, Cambridge, USA
}

\renewcommand{\arraystretch}{2} % for table

\graphicspath{{./images/}}

\begin{document}
\title{Efficient Protein Ground State Energy Computation via Fragmentation and Reassembly}

\author{Laia Coronas Sala}
\affiliation{\LDIGEU}
\email{laia.coronas@lighthouse-dig.com}

%\author{Guillermo Alonso-Linaje}
%\affiliation{\UVA}
%\email{guillermo@alonso-linaje.com}

\author{Parfait Atchade-Adelomou}
%\affiliation{\DSD}
%\email{parfait.atchade@salle.url.edu}
\affiliation{\LDIGEU}
\affiliation{\LDIG}
\email{parfait.atchade@lighthouse-dig.com}
\affiliation{\MIT}
\email{parfait@mit.edu}

\date{Jan 2024}

\begin{abstract}
Protein characterization is one of the key components for understanding the human body and advancing drug discovery processes. While the future of quantum hardware holds the potential to accurately characterize these molecules, current efforts focus on developing strategies to fragment larger molecules into computationally manageable subsystems. In this work, we propose a novel strategy to enable quantum simulation using existing quantum algorithms. Our approach involves fragmenting proteins into their corresponding amino acids, simulating them independently, and then reassembling them post-simulation while applying chemical corrections. This methodology demonstrates its accuracy by calculating the ground state energy of relatively small peptides through reassembling, achieving a mean relative error of only $0.00469 \pm 0.01071\%$. Future directions include investigating, with larger quantum computers, whether this approach remains valid for larger proteins.

%\newline
%\newline
\end{abstract}

\maketitle

\section{Introduction}

Quantum mechanics delves into particle behavior at subatomic scales, capturing interactions and energy states with unparalleled precision \cite{baiardi2023quantum}. Unlike homology-dependent classical methods, quantum chemistry is rooted in the physical principles of energy and entropy, making it ideal for high-precision \textit{ab initio} conformational predictions \cite{vanmourik2004firstprinciples, bryce2020next}. These predictions are extremely valuable in the context of peptides, as they are principal targets in drug discovery processes \cite{usman2021peptides}.

In this context, \textit{Quantum Computing} (QC) emerges as a promising tool to overcome computational barriers, particularly in the calculation of \textit{Ground State Energy} (GSE), a property that provides crucial insights into molecular stability, reactivity, and interaction dynamics \cite{britt2004, Moll2018, nielsen2010quantum}. This property can be assessed through various quantum and classical strategies, such as the \textit{Variational Quantum Eigensolver} (VQE), \textit{Quantum Phase Estimation} (QPE), or classical approximations like \textit{Hartree-Fock} (HF) \cite{kirby2021, QPE, Matveeva2023}.

However, given the current limitations of quantum hardware \cite{nielsen2010quantum,atchadeadelomou2022quantum}—such as noise, scalability issues, and inadequate error correction—simulating large and complex molecules remains a significant challenge \cite{blunt2022}. To address these constraints, we propose a hybrid strategy that leverages both classical and quantum techniques, suitable for the \textit{intermediate-scale quantum} (ISQ) era.

In this approach, peptides are fragmented into smaller, more manageable components represented by their constituent amino acids. These amino acids provide a fundamental chemical and biological framework, simplifying the computational process. Then, the GSE of each individual amino acids is computed to later reassemble the molecule and study the feasibility of this approach.

This paper is organized as follows: Section~\ref{sec:motivation} outlines the motivations, hypotheses, and assumptions underpinning the proposed methodology's design and development. Section~\ref{sec:state_art} reviews the state of the art. Section~\ref{sec:background} provides background information on proteins and amino acid bonds. Section~\ref{sec:methodology} describes the proposed methodology in detail. Section~\ref{sec:validation} explains the validation process employed. Section~\ref{sec:results} presents and analyzes the experimental results, while Section~\ref{sec:discussion} offers a critical discussion. Finally, Section~\ref{sec:conclusions} concludes by summarizing the implications of the work and suggesting directions for future research.

\section{Motivation and Hypotheses}\label{sec:motivation}

The objective from a biological standpoint is to address the challenge of efficiently simulating proteins by fragmenting them into amino acids. Proteins are the fundamental basis for all interactions in the body, and their characterization is essential for understanding biological processes and advancing drug discovery \cite{nelson2021lehninger, alberts2015molecular}. Notably, all proteins in the human body are composed of the same $20$ amino acids \cite{voet2011biochemistry}. Therefore, we aim to develop a method for fragmenting proteins into their corresponding amino acids and then reassembling them post-simulation, incorporating chemical corrections to ensure the accurate reconstitution of the original molecule.

This work is therefore based on a set of hypotheses and assumptions that guide the design and development of the proposed methodology. First, it is assumed that large molecules, such as proteins, can be fragmented into amino acids without significantly compromising the global properties of the molecule. This assumption is supported by the fact that amino acids are chemically well-defined and biologically relevant units that can be studied independently and reassembled. Second, it is further assumed that any residual interactions between fragments can be addressed through specific energetic adjustments. Third and last, it is acknowledged that current quantum computers are noisy and limited in terms of available qubits, so the proposed methodology is designed to ensure the efficient implementation of the strategy.

\section{State of the Art}\label{sec:state_art}

Molecular fragmentation is based on the principle of simplifying a complex molecular system by breaking it down into smaller, more manageable parts while retaining the essential chemical interactions that govern its behavior \cite{collins2006}. However, many fragmentation methods lack a unified framework to ensure accuracy during the recombination of the fragments, which can introduce errors in the final energy calculations. Previous studies have generally shown success with small, non-polar, and less charged molecules but struggle to generalize to larger, more complex, and highly correlated systems \cite{li2005, deev2005, bettens2006}.

After fragmentation, the first step involves simulating and computing the energies of each fragment individually. In our previous work, we explored the application of VQE and QPE to estimate GSE \cite{coronas2024leveraging}. Our experiments revealed that while both methods showed promise for small systems, their performance decreased as the system size increased, highlighting the importance of effective fragmentation. Specifically, VQE faced scalability challenges as molecular complexity grew, limiting its applicability to larger systems, as corroborated by prior results \cite{Moll2018, kirby2021, peruzzo2014variational, kandala2017hardware}. On the other hand, with QPE we can select the lowest energy (eigenvalue), achieving higher precisions \cite{Kang2022, Russo2021, QPE}. This observation motivated a focus on approaches that integrate QPE with advanced quantum algorithms to improve initial state preparation and enhance simulation efficiency \cite{dcunha2024state, choi2024probing}. Nevertheless, very good accuracies were also achieved using the HF approximation, a classical method that proved easier to implement and computationally efficient.

Following fragmentation and simulation, the final step is reassembling the information from each fragment to reliably represent the original molecule. A standard method for this is the energy-based fragmentation approach, where the total GSE is derived from summing the energies of the embedded subsystems \cite{Li2007}. However, this approach often fails to account for necessary chemical corrections that could improve accuracy. Common reassembly techniques, such as simulating interactions between fragments based on the proximity of radicals in peptide chains, attempt to address these corrections by modeling interactions as fragments come closer in space \cite{collins2006, li2005, Li2007}. Despite their utility, these methods can become complex and less reliable for long peptide chains.

A review of the current state of the art highlights the absence of a standardized pipeline for reliably estimating the GSE of larger molecules, such as peptides. Existing approaches often face challenges in balancing computational efficiency with accuracy, particularly when quantum algorithms are applied to larger fragments. To overcome these limitations, we propose a strategy that emphasizes a systematic division and reintegration of fragments, complemented by chemical corrections to enhance accuracy. 

\section{Theoretical Background} \label{sec:background}

Proteins are essential macromolecules in biological systems, responsible for a wide variety of functions such as catalyzing reactions, transmitting signals, providing structure, and facilitating movement \cite{usman2021peptides}. These proteins are composed of long chains of amino acids, which are the basic building blocks of these molecules. The sequence and composition of these amino acids determine the protein’s structure and function. Each amino acid has a common structure consisting of a \textit{central carbon atom} (\(C_\alpha\)) bonded to an \textit{amino group} (-NH\(_2\)), a \textit{carboxyl group} (-COOH), a \textit{hydrogen atom} (-H), and a \textit{radical group} (side chain). This side chain varies among the 20 amino acids and defines their chemical properties, playing a critical role in determining how the protein folds and interacts with other molecules \cite{alberts2015molecular, voet2011biochemistry, clayden2001organic}.

\begin{figure}[ht]
    \centering
    \includegraphics[width=0.9\linewidth]{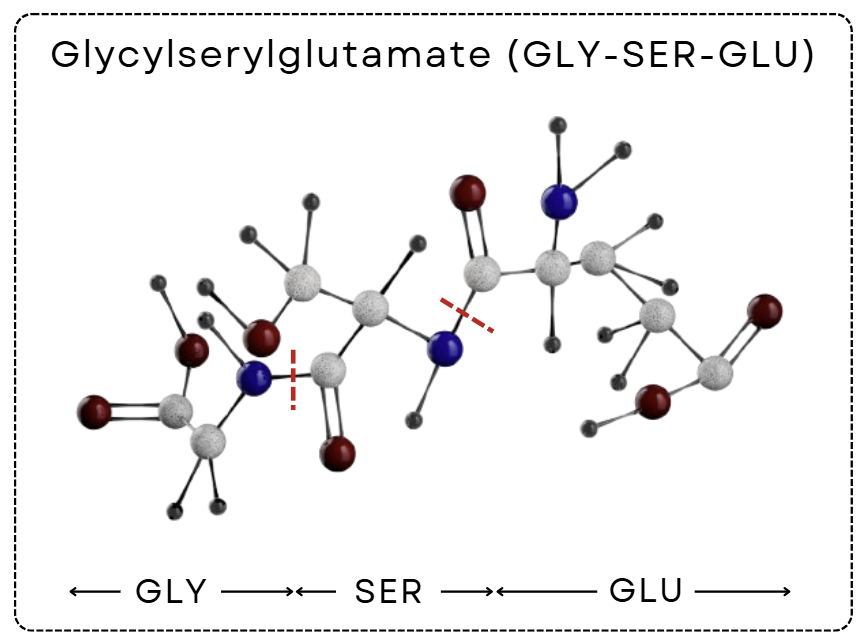}
    \caption{Structure of a small peptide composed of three amino acids: glycine (GLY), serine (SER), and glutamic acid (GLU). In red, the division clearly showing the different amino acids and the peptide bonds uniting them, which will be our fragmentation point.}
    \label{fig:glut}
\end{figure}

To form a protein, amino acids are linked together by a covalent bond known as the peptide bond, which forms through a dehydration synthesis (condensation) reaction. In this process, the carboxyl group (-COOH) of one amino acid reacts with the amino group (-NH\(_2\)) of another amino acid, resulting in the release of a \textit{water molecule} (H\(_2\)O). This loss of a water molecule is a key characteristic of peptide bond formation, and we strongly believe it should be taken into account for GSE calculations, as we will present later \cite{clayden2001organic}.  

In computational chemistry and molecular simulations, it is often useful to fragment proteins at the peptide bond to simplify the modeling of large molecules \cite{collins2006}. The peptide bond serves as a natural and stable point for fragmentation because it is a well-defined, relatively strong bond. By cutting the protein at the peptide bond, smaller peptide fragments can be isolated and studied independently, reducing the computational complexity. The main limitation, however, lies in how to reintegrate the information obtained from each fragmented amino acid \cite{collins2006}. Figure \ref{fig:glut} provides an illustrative example of a small peptide composed of three amino acids: glycine, serine, and glutamic acid.

\section{Proposed Methodology} \label{sec:methodology}
\subsection{General Pipeline}

To address the objectives presented, we first propose a general pipeline for the quantum simulation of any protein. This process should begin with the fragmentation of the protein into its constituent amino acids. Each amino acid is then prepared for simulation to compute its GSE. If, however, the amino acid is still too large for simulation (depending on the composition of its side chain), we propose further fragmentation. Additionally, to accurately simulate each fragment, we suggest combining initial state preparation with QPE, to ensure efficiency. Finally, the last step is to reintegrate the energies of each amino acid, taking into account chemical corrections based on the initial protein structure.

\begin{figure}[ht]
    \centering
    \includegraphics[width=1\linewidth]{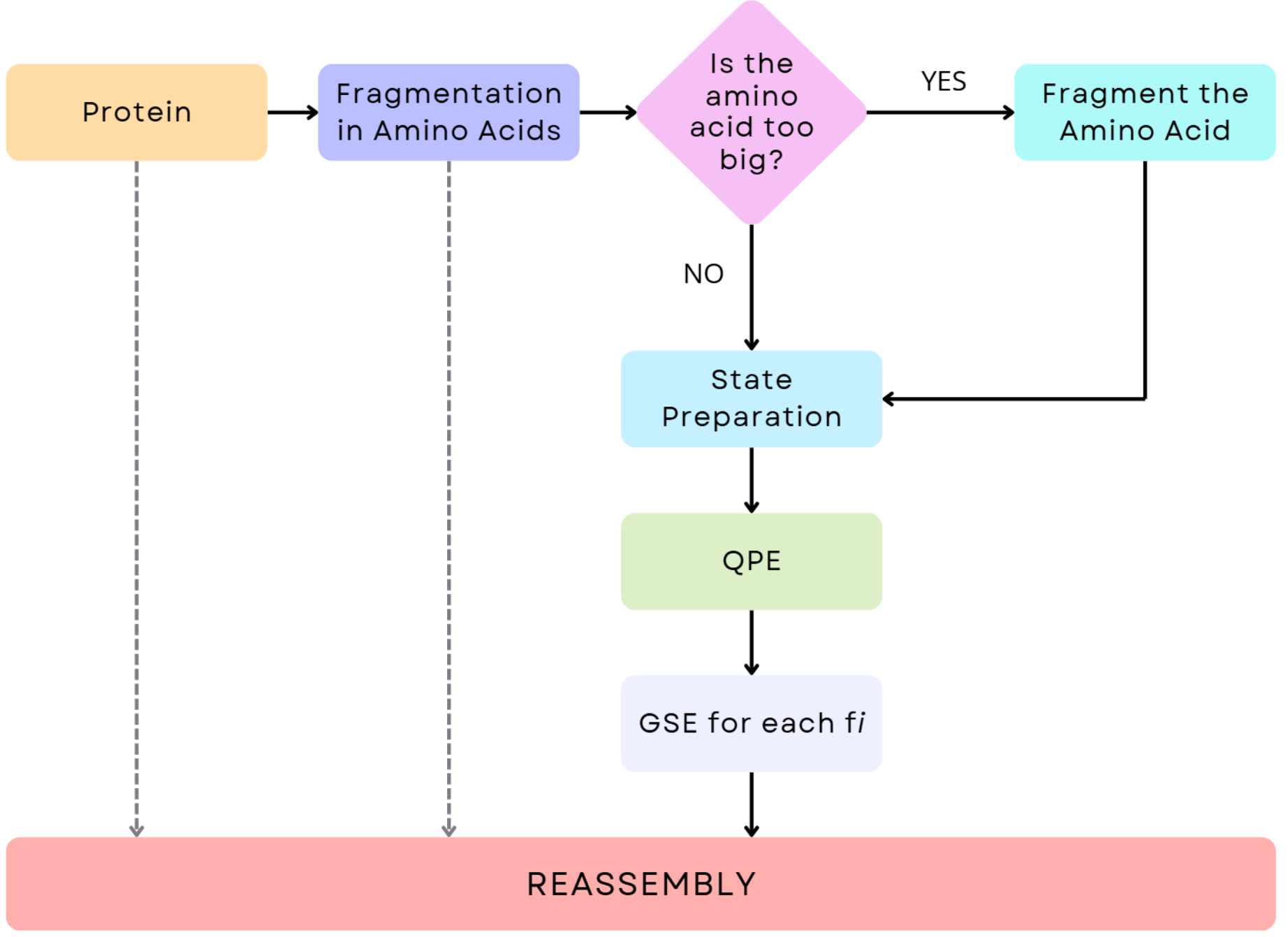}
    \caption{General pipeline proposed for the characterization of any possible protein. GSE stands for Ground State Energy, QPE for Quantum Phase Estimation, and $f_i$ corresponds to any possible fragment generated from the original molecule.}
    \label{fig:2}
\end{figure}

\subsection{Molecular Fragmentation, Reassembly and Energy Adjustments}

The first step involves dividing large molecules into smaller, more manageable fragments. To this end, we identified 20 amino acids present in all proteins in the human body and generated a data structure that facilitated their analysis and future integration into molecular simulation applications. After calculating the fragment energies, the results were combined to estimate the total energy of the original molecule. This was achieved by directly adding the energy estimated for each of the fragments, in this case, for each amino acid present in the corresponding peptide. Additionally, energy adjustments were applied based on protein formation and composition to improve the accuracy of the final result. Overall, the reassembling of the energies from each fragment could be expressed by the following equation:

\begin{equation}
    E_{m} = \sum_{i=1}^{n} E_{f_i} \pm \sum_{j=1}^{k} E_{am_j},
    \label{eq:fragment_str}
\end{equation}

where \(n\) is the protein's total number of fragments generated, \(E_{f_i}\) is the GSE corresponding to each fragment, \(k\) is the total number of molecules lost or added within the process, \(E_{am_j}\) corresponds to the GSE of each additional molecule, and \(E_m\) corresponds to the final GSE of the molecule obtained through reassembly. Here, \(i\) is the index that enumerates the \(n\) fragments of the protein generated, with \(i = 1, 2, \dots, n\). Similarly, \(j\) represents the index of the \(k\) additional or removed molecules during the process, with \(j = 1, 2, \dots, k\).

\begin{figure}[ht]
    \centering
\includegraphics[width=1\linewidth]{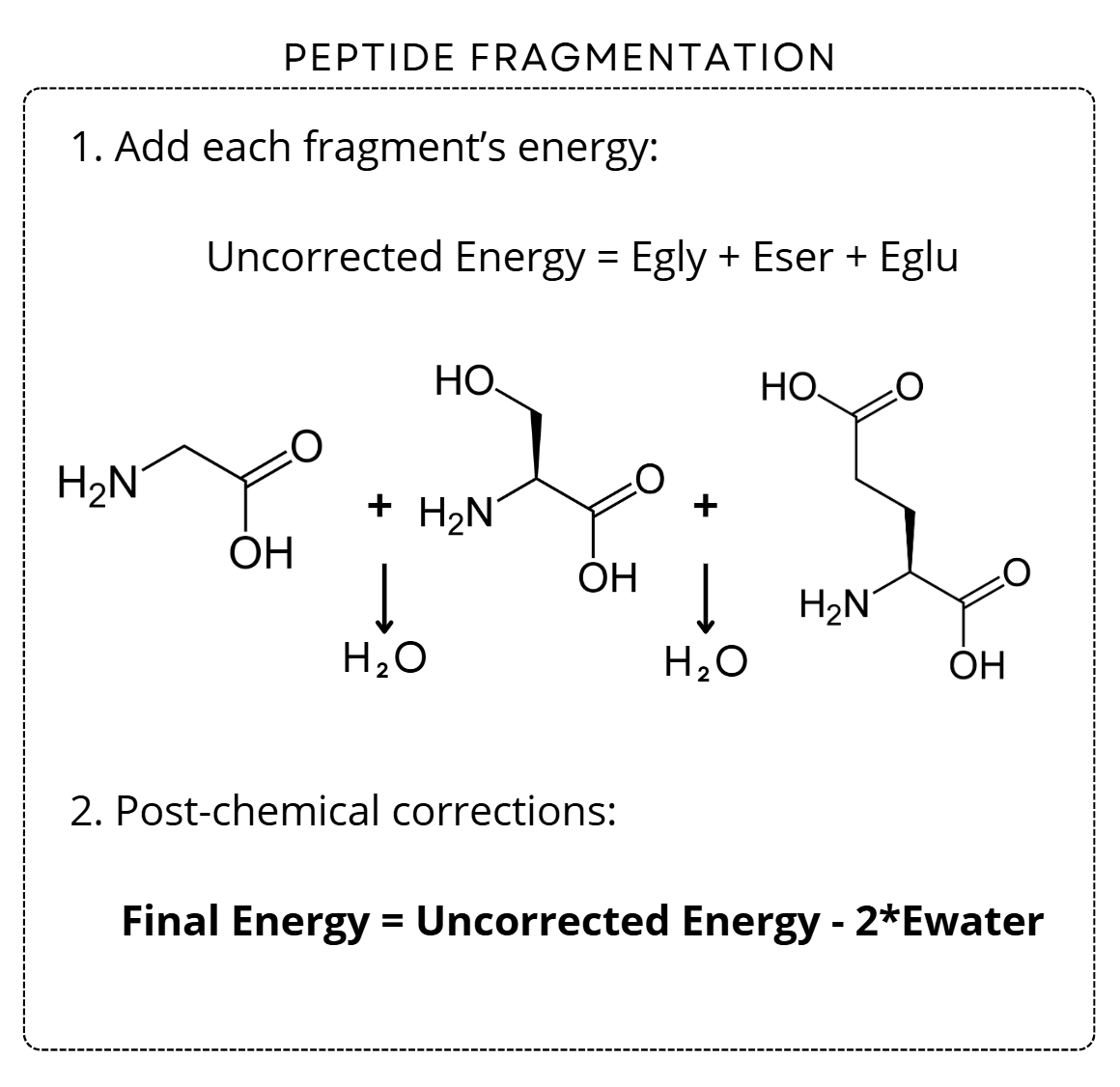}
    \caption{Peptide fragmentation and reassembly. We divided the molecule presented in Figure 1 into its corresponding amino acids and then calculated each GSE independently. We then added them all together and subtracted the energy corresponding to the two water molecules lost during the peptide bonds condensation to form the peptide.}
    \label{fig:process_1}
\end{figure}

This means that we divided the protein into its constituent amino acids, calculated the energy for each amino acid, and then summed each energy to obtain the total energy. Afterwards, we subtracted the energy corresponding to the molecules lost through bond formation to obtain the final GSE. For instance, if the molecule lost was a water molecule, which is typically lost in peptide bond formation, we subtracted $-74.97$ Ha from the final energy (the GSE of \ce{H2O} \cite{ccbdb}). If the molecule lost was another one, we computed the GSE of that molecule to also subtract it. If the molecule had any modifications, for instance, an added functional group, we also added the GSE corresponding to that group alone (as we will see with the Aspartame peptide). With this, we obtained the final reassembled GSE of the protein. Figure \ref{fig:process_1} shows the process for the protein presented in Figure \ref{fig:glut}.

\begin{figure}[ht]
    \centering
\includegraphics[width=0.8\linewidth]{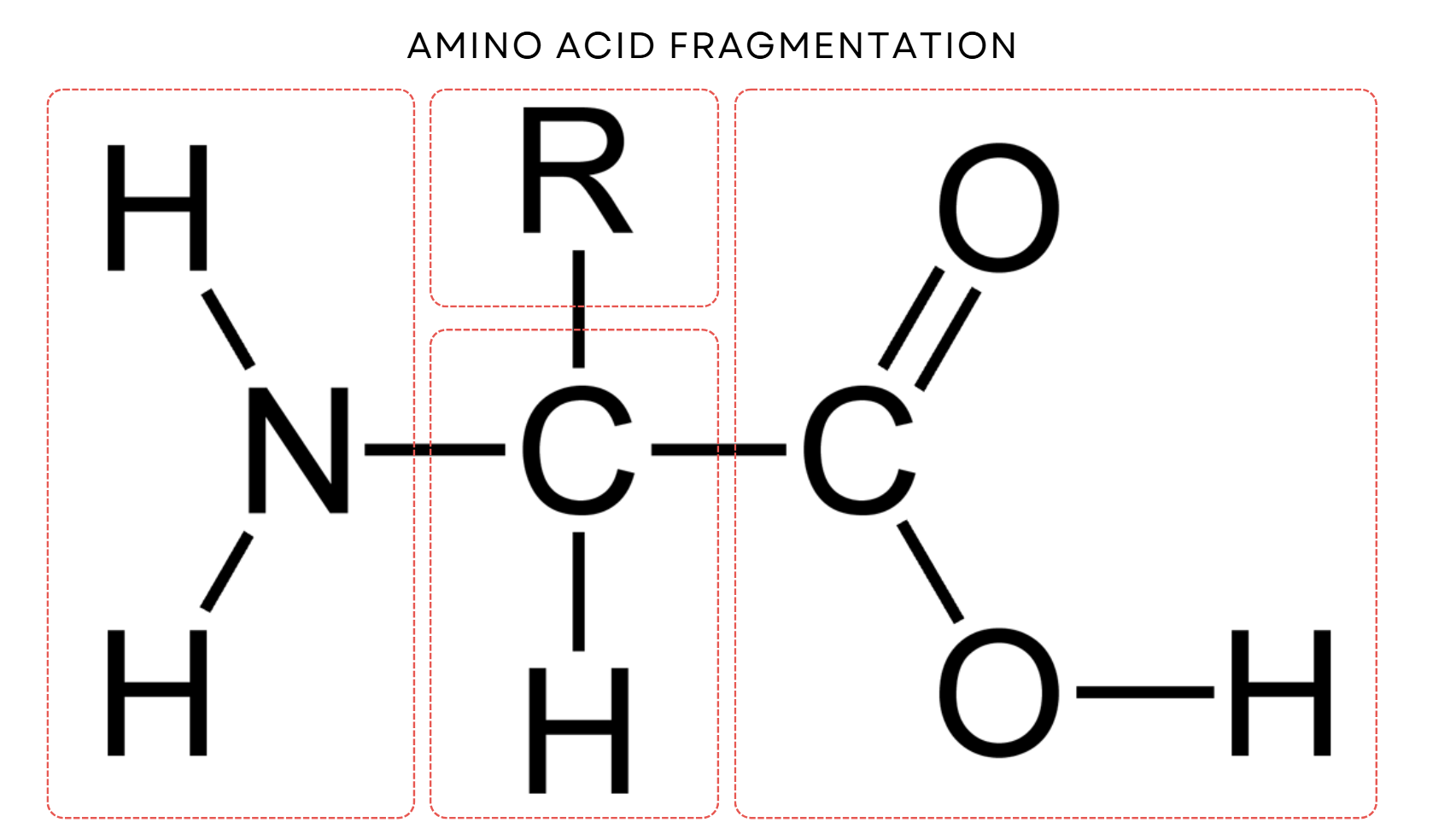}
    \caption{Amino acid fragmentation. We divided the molecule  into the carboxyl group, the amino group, the side chain (R) and the remaining \ce{CH} group, and calculated each GSE independently. We then added them all together without further correction.}
    \label{fig:aa}
\end{figure}

However, it is important to acknowledge the significant challenges inherent in performing quantum simulations of individual amino acids. For example, simulating \textit{glycine}, the smallest amino acid, requires the use of $60$ qubits and involves more than $709,671$ terms in its Hamiltonian representation within the \textit{STO-3G} basis. Therefore, depending on the available computational resources, we propose further fragmentation of individual amino acids using the same reassembly approach. Specifically, the amino acid is divided into four distinct groups: \ce{-NH2}, \ce{-COOH}, \ce{-CH}, and the corresponding side chain (R). The energy of each fragment would then be computed and reassembled according to Equation \ref{eq:fragment_str}, by summing the energy of each fragment without applying additional corrections, as no water molecules are lost during the bond formation process. Nevertheless, it should be noted that further fragmentation may introduce slightly higher errors. Figure ,\ref{fig:aa} shows the fragmentation at an amino acid level.

\section{Validation}\label{sec:validation}

To evaluate the feasibility of the fragmentation and reassembly approach, we tested our methodology on a set of 20  relatively small peptides with diverse chemical and structural properties, and on the 20 basic amino acids. This allowed us to validate the accuracy of the approach and ensure its applicability across different molecular configurations. Energy calculations were performed using the well-established Hartree-Fock (HF) methodology, leveraging the precision achievable with today’s classical computing capabilities, ensuring a robust baseline independent of the potential advancements that future quantum computers might bring.

The 3D atomic coordinates and molecular system types were extracted from PubChem SDF files \cite{pubchem} and processed using the PySCF package for HF calculations \cite{pyscf_scf}. Specifically, we employed the restricted \textit{Hartree-Fock} (RHF) method with a minimal basis set (STO-3G) to calculate the total energy of the system. All calculations were performed on a high performance set-up featuring a 13th Gen Intel(R) Core(TM) i7-13700H processor, 32 GB of RAM, and a 64-bit operating system running on the \textit{Windows Subsystem for Linux} (WSL).

To evaluate the performance of our method, we utilized the \textit{Mean Percentage Relative Error} (RE\%) and its \textit{Standard Deviation} (STD) (Equation \ref{eq:relative_err}). Specifically, we calculated the percentage error between the reassembled energies and the classically computed energies of the entire molecule, which we treated as the \textit{Ground Truth} (GT).

\begin{equation}
    \mu_{\text{RE (\%)}} = \frac{1}{n} \sum_{i=1}^{n} \left| \frac{E_{m_{i}} - \text{GT}_i}{\text{GT}_i} \right| \times 100.
    \label{eq:relative_err}
\end{equation}

Where \(n\) is the total number of data points, and $E_{m_{i}}$ and \(\text{GT}_{i}\) represent the reassembled and the GT GSE for the \(i\)-th peptide, respectively.

\section{Results}\label{sec:results}

Table \ref{tab:results} presents a comparison between the GT GSE and the GSE obtained through reassembly for different small peptides, fragmenting the peptides only in their corresponding entire amino acids. The results demonstrate that the reassembled GSE values closely align with the GT values, showing minimal deviations across all peptides. Notably, this accuracy is maintained even in cases of uncommon bond formations, such as aspartame (involving the addition of a methyl group) and cystine (involving the loss of \ce{H2} instead of a water molecule). The calculated mean \%RE is $0.00469\%$ with a STD of $0.01071\%$, highlighting the good accuracy of our approach. 

\begin{table}[ht]
    \centering
    \renewcommand{\arraystretch}{1.2}
    \captionsetup{skip=10pt}
    \caption{Results obtained for the strategy presented. The first column corresponds to the peptide simulated, the second column to the GT GSE computed classically, the third column to the GSE obtained through the reassembling of each fragment's energy, and the fourth column to the RE\% between the second and third columns.}
    \label{tab:results}
    \begin{tabular}{|l|r|r|r|}
        \hline
        \textbf{Peptides} & \textbf{GT GSE [Ha]} & \textbf{Em [Ha]} & \textbf{RE [\%]} \\ \hline
        Gly-Gly            & -483.23779          & -483.25713       & 0.00400        \\ \hline
        Gly-Ala            & -521.82046          & -521.83697       & 0.00317        \\ \hline
        Gly-Ser            & -595.64593          & -595.66156       & 0.00262        \\ \hline
        Carnosine (Ala-His) & -781.24420         & -781.25935       & 0.00194         \\ \hline
        Aspartame (Asp-Phe) & -1010.80954        & -1011.31538      & 0.05004         \\ \hline
        Cystine (Cys-Cys)   & -1420.58943        & -1420.59711      & 0.00054       \\ \hline
        Leu-Thr            & -788.53834          & -788.55273       & 0.00182         \\ \hline
        Thr-Lys            & -842.85520          & -842.87108       & 0.00188         \\ \hline
        Trp-His            & -1137.12083         & -1137.14177      & 0.00184         \\ \hline
        Phe-Ile            & -902.88517          & -902.89959       & 0.00160        \\ \hline
        Arg-Met            & -1308.21468         & -1308.22803      & 0.00102      \\ \hline
        Ser-Cys            & -1027.39113         & -1027.40735      & 0.00158         \\ \hline
        Tyr-Asp            & -1046.06184         & -1046.07719      & 0.00147        \\ \hline
        Glu-Gly            & -745.47664          & -745.49542       & 0.00252         \\ \hline
        His-Arg-Val        & -1378.59438         & -1378.62213      & 0.00201         \\ \hline
        Val-Asp-Ser        & -1139.16790         & -1139.19877      & 0.00271         \\ \hline
        Gly-His-Lys        & -1155.42028         & -1155.45421      & 0.00294         \\ \hline
        Val-Ala-Ser        & -954.09354          & -954.12300       & 0.00309         \\ \hline
        Gly-Val-Ala        & -841.68835          & -841.71857       & 0.00359         \\ \hline
        Ser-Gly-Glu        & -1062.00937         & -1062.04546      & 0.00340       \\ \hline
    \end{tabular}
\end{table}

Table \ref{tab:results_aa} presents the results accounting for the fragmentation of the amino acids into their corresponding four different groups presented in Figure \ref{fig:process_1}. In this case, the error is higher due to the further fragmentation of the molecules but still very promising (mean \%RE of $0.26844\% \pm 0.32794\%$).

\begin{table}[ht]
    \centering
    \renewcommand{\arraystretch}{1.2}
    \captionsetup{skip=10pt}
    \caption{Results obtained for the energy reassembly strategy. The first column corresponds to the amino acid simulated, the second column to the total energy computed directly (GT), the third column to the energy obtained through reassembly, and the fourth column to the error percentage.}
    \label{tab:results_aa}
    \begin{tabular}{|l|r|r|r|}
        \hline
        \textbf{Amino Acid} & \textbf{GT GSE [Ha]} & \textbf{Em [Ha]} & \textbf{RE [\%]} \\ \hline
        Histidine       & -538.53389  & -537.58932  & 0.17540 \\ \hline
        Leucine         & -433.42225  & -434.01055  & 0.13573 \\ \hline
        Isoleucine      & -433.42805  & -434.01055  & 0.13439 \\ \hline
        Lysine          & -487.74061  & -487.36827  & 0.076339 \\ \hline
        Methionine      & -788.02139  & -787.09064  & 0.11811 \\ \hline
        Phenylalanine   & -544.43743  & -544.04328  & 0.072395 \\ \hline
        Threonine       & -430.09637  & -429.12416  & 0.22604 \\ \hline
        Tryptophan      & -673.57378  & -673.15017  & 0.062891 \\ \hline
        Valine          & -394.84750  & -394.45688  & 0.098928 \\ \hline
        Arginine        & -595.17255  & -594.18510  & 0.16591 \\ \hline
        Cysteine        & -710.85730  & -715.26885  & 0.62060 \\ \hline
        Glutamine       & -521.82179  & -516.80354  & 0.96168 \\ \hline
        Asparagine      & -483.23923  & -479.61163  & 0.75068 \\ \hline
        Tyrosine        & -618.27595  & -611.07066  & 1.1654 \\ \hline
        Serine          & -391.51594  & -391.12360  & 0.10021 \\ \hline
        Glycine         & -279.11151  & -278.67407  & 0.15673 \\ \hline
        Aspartic acid   & -502.76713  & -502.31178  & 0.090569 \\ \hline
        Glutamic acid   & -541.34980  & -540.89235  & 0.084502 \\ \hline
        Proline         & -393.70020  & -393.87365  & 0.044055 \\ \hline
        Alanine         & -317.69136  & -317.28420  & 0.12816 \\ \hline
    \end{tabular}
\end{table}

\section{Discussion}\label{sec:discussion}

The primary objective of this work was to establish an efficient pipeline for computing the GSE of proteins. We hypothesized that proteins could be fragmented into amino acids, each fragment could be simulated individually, and the energies could then be reassembled by applying chemical corrections associated with the formation of the original molecule.

Our results were very promising, with a calculated \%RE of $0.00469 \pm 0.01071\%$ for the fragmentation in entire amino acids and a \%RE of $0.26844\% \pm 0.32794\%$ for amino acid fragmentation. Hence, computing energies by having to fragment both the protein and the amino acid would mean potentially assuming a \%RE of $0.27313\%$. This suggests that our reassembling approach delivers highly accurate results, with minimal deviations from the true GSE values, highlighting its potential for future applications in molecular simulations and quantum chemistry. 

The presented approach involved summing the energy of each amino acid and adding corrections by incorporating or subtracting molecules involved in the formation of the original peptide. Remarkably, this method proved to be effective even in cases involving uncommon bond formations. For example, cystine, a dimer formed by two cysteines, is typically bonded through a sulfur link instead of the \textit{carboxyl} and amino terminal groups. In such cases, we subtracted the GSE of two hydrogen molecules (\(\ce{H2}\)) rather than a water molecule during the reassembly process, still obtaining highly accurate results. Similarly, in the case of aspartame, a dimer consisting of aspartic acid and phenylalanine, a methyl group (\(\ce{CH3}\)) is added during the formation of the molecule. To account for this, we computed the GSE of \(\ce{CH3}\) and added it to the final energy, again achieving very high accuracy. This is of utmost importance since, unlike previous methodologies, we provide a very easy pipeline to implement \cite{collins2006}.

Similarly, the accuracy was almost maintained even for amino acid fragmentation, where molecules were completely divided into much smaller fragments. Despite this, our methodology appears to effectively reconstitute the properties of the original molecule without significant losses. Interestingly, this was true for almost all side chains of amino acids, even though some were much more complex than others. 

Despite its effectiveness for simulating small fragments, the HF approximation has limitations, particularly for larger fragments, due to its reliance on classical approximations. Until QC advances allow for more efficient simulations, HF or its classical alternative, like density matrix renormalization group (DMRG), remains a practical approach to address current constraints in quantum simulations \cite{baiardi2023quantum, schollwock2005density}. Future work will focus on integrating advanced quantum methods such as the EVA technique \cite{alonsolinaje2021evaquantumexponentialvalue}, QPE, QPE derivatives \cite{loaiza2024nonlinear}, and efficient initial state preparation, which have demonstrated their effectiveness in calculating GSE \cite{coronas2024leveraging}. Additionally, as even individual amino acids often require a significant number of qubits for accurate simulation, future research should explore further fragmenting amino acids while addressing the challenges of higher error rates.
Lastly, while the results obtained for small peptides are promising, further investigation is required to evaluate the scalability of this approach for larger proteins. Our experiments were conducted on a set of twenty relatively small peptides, but extending the methodology to larger molecules may introduce additional computational complexities and challenges.

\section{Conclusions}\label{sec:conclusions}

We proposed a novel approach for quantum protein simulations by fragmenting proteins into their constituent amino acids, simulating each independently, and reassembling them post-simulation with chemical corrections. Additionally, we introduced further fragmentation of amino acids into computationally manageable groups. Our strategy addresses current computational limitations in simulating large molecular systems, combining classical and quantum methods to achieve both efficient and accurate results. The validation of our method, performed through the calculation of the ground-state energy for small peptides, demonstrated that the reassembled energies closely match the ground truth, with a relative error of only $0.00469 \pm 0.01071\%$ for individual amino acid fragmentation and  $0.26844\% \pm 0.32794\%$ when further fragmentation was applied. These results highlight the high accuracy of our approach and its potential for application to larger and more complex protein systems. Future work should focus on integrating advanced quantum techniques, such as initial state preparation and quantum phase estimation, to improve the accuracy and efficiency of the simulations. Moreover, this approach should be tested with larger proteins to validate the scalability of the proposed strategy.
\vspace{3mm}
\section*{Code}
The code to reproduce the figures and explore additional settings is available in the following GitHub repository: \href{https://github.com/laiacoronas/protein_fragmentation}{https://github.com/laiacoronas/protein\_fragmentation}.

\section*{Acknowledgements} The authors thank Guillermo Alonso-Linaje for the discussions and consideration during the experiments. The authors also thank Mireia Alcón Cans for her contributions to the figures in this paper.

\bibliographystyle{unsrt}
\bibliography{references}
\end{document}